\documentclass[prl,showpacs,hyperref,superscriptaddress,twocolumn]{revtex4-1}
\usepackage{amsfonts}
\usepackage{amsmath}
\usepackage{amssymb}
\usepackage{color}

\usepackage{graphicx}
\usepackage{pdfpages}
\usepackage{epsfig}
\usepackage{subfigure}

\begin{document}

\title{Spin squeezing in an ensemble of quadrupolar nuclei NMR system}

\author{R. Auccaise} \email{raestrada@uepg.br}
\affiliation{
Departamento de F\'{\i}sica, Universidade Estadual de Ponta Grossa, Av. Carlos Cavalcanti, 4748, 84030-900 Ponta Grossa, Paran\'{a}, Brazil}

\author{A. G. Araujo-Ferreira}
\affiliation{
Instituto de F\'{\i}sica de S\~{a}o Carlos, Universidade de S\~{a}o Paulo, Caixa Postal 369, 13560-970 S\~{a}o Carlos, S\~{a}o Paulo, Brazil}

\author{R. S. Sarthour}
\affiliation{
Centro Brasileiro de Pesquisas F\'{\i}sicas, Rua Dr. Xavier Sigaud 150, 22290-180 Rio de Janeiro, Rio de Janeiro, Brazil}

\author{I. S. Oliveira}
\affiliation{
Centro Brasileiro de Pesquisas F\'{\i}sicas, Rua Dr. Xavier Sigaud 150, 22290-180 Rio de Janeiro, Rio de Janeiro, Brazil}

\author{T. J. Bonagamba}
\affiliation{
Instituto de F\'{\i}sica de S\~{a}o Carlos, Universidade de S\~{a}o Paulo, Caixa Postal 369, 13560-970 S\~{a}o Carlos, S\~{a}o Paulo, Brazil}

\author{I. Roditi} \email{roditi@cbpf.br}
\affiliation{
Centro Brasileiro de Pesquisas F\'{\i}sicas, Rua Dr. Xavier Sigaud 150, 22290-180 Rio de Janeiro, Rio de Janeiro, Brazil}

\begin{abstract}
We have characterized spin-squeezed states produced at a temperature of $26^\circ{\mathrm C}$ on a Nuclear Magnetic Resonance (NMR) quadrupolar system. The implementation is carried out in an ensemble of $^{133}$Cs nuclei with spin $I=7/2$ of a lyotropic liquid crystal sample. We identify the source of spin squeezing due to the interaction between the  quadrupole moment of the nuclei and the electric field gradients internally present in the molecules. We use the spin angular momentum representation to describe formally the nonlinear operators that produce the spin squeezing. The quantitative and qualitatively characterization of the spin squeezing phenomena is performed through a squeezing parameter and squeezing angle developed for the two-mode BEC system, and, as well, by the Wigner quasi-probability distribution function. The generality of the present experimental scheme indicates its potential applications on solid state physics.
\end{abstract}

\pacs{03.65.Wj, 76.60.-k, 42.50.Dv, 03.67.-a}
\maketitle  

The role of non-classical states in atomic physics and optics has been extensively investigated in the last two decades \cite{nori,dellanno,gross2012}. A great deal of this interest is due to the collective behavior of atoms in the so called - Spin-Squeezed State.  Quantum states of this kind were first studied in relation to the production of atom-atom entanglement using a non-linear spin-spin interaction \cite{kitagawa1991, kitagawa1993} - the one-axis twisting (OAT) model. Since then, many theoretical developments \cite
{wineland1992,duan2000,sorensen2001,helmerson2001,takeuchi2005,law2001,jin2007PRA,jin2007PRL,boixo2008,rudner2011,diaz2012} led to various applications in the domain of quantum information processing including, for instance, proposals that spin squeezing may be exploited in quantum entanglement \cite{sorensen2001,korbicz2005,he2011}, also in quantum metrology, where one explores the idea of a fundamental noise limit set by quantum mechanics laws \cite{gross2010,boixo2008}, as well as in atom chip based investigations \cite{riedel2010}.

With respect to experimental observations \cite
{hald1999,chaudhury2007,jo2007,sebby-strabley2007,fernholz2008,takano2009,esteve2008,gross2010,riedel2010},
spin-squeezed states have been attained mainly in many-body atom-light interaction scenarios by use of the collective spin in a sample of $N$ atoms \cite{kuzmich2000,takano2009}.  These kind of states were also seen by exploring the internal structure (nuclear plus electronic spin) of individual atoms for the hyperfine ground states $F=3$ and $F=4$ \cite{chaudhury2007,fernholz2008} in a Hilbert space of dimension $(2F+1)=7$ and $9$.

Therefore, as one can recognize from the above, many theoretical and experimental efforts have been directed to this compelling phenomena within condensed matter physics. Nevertheless, in  the domain of soft matter physics, as well as in solid state physics, there is significantly less activity. To the extent of our knowledge, only within the context of quantum simulation \cite{sinha2003} one experimental implementation was performed discussing a squeezing process for the case of liquid state Nuclear Magnetic Resonance (NMR). This absence of results in soft and solid matter physics may be understood by two main challenges: heterogeneous coupling values between spin particles, that produce different free evolutions; and high temperature regime, that rapidly destroys their quantum coherence. Basically, the ability of controlling the couplings between particles at the quantum level, and collective behaviors of molecules to compensate thermal vibrations play a key role in order to execute efficiently an experimental implementation in those physical systems. Recently, these requirements were fulfilled in a liquid crystal platform within the NMR framework, carrying out experimental tasks that apply the concept of coherent states, one- and two-mode BEC like system, and protocols of classical bifurcation \cite{auccaise2013,auccaise2009,araujo-ferreira2013}.  Such approaches match a low dimensionality regime as analogously happens in many body systems \citep{chaudhury2007,fernholz2008}.

These experimental and theoretical results led us to a novel discussion of the
spin squeezing phenomena in NMR. In many body investigations, part of the theoretical framework comes from studies using the two-mode BEC system \cite{law2001,jin2007PRA,jin2007PRL,diaz2012}. In these, in order to reach a spin-squeezed state, an OAT model is implemented. The associated Hamiltonian is also known as the two-site Bose-Hubbard model and the particle description can be mapped into an angular momentum description using the Schwinger representation. More specifically, it is possible to map a Hamiltonian described in terms of creation $\hat{a}^{\dagger}_i$ and annihilation $\hat{a}_i$ operators, satisfying the commutation relations $[\hat{a}_i,\hat{a}^{\dagger}_j]=\delta_{ij}$ ($i,j=1,2$), to an angular momentum algebra representation  \cite{nori,gross2012,wineland1992,diaz2012,milburn1997,law2001,jin2007PRL}.
Another related discussion about squeezing  has been carried out in quantum dots system, and is due to the nuclear-electron interaction \cite{rudner2011}. The common formalism used in these investigations is the collective spin representation. Now, the relevance of those results for our case is that it is possible to explore the same formalism in the context of an ensemble of NMR quadrupolar nuclei. On that account, here we report the experimental observation of the dynamics of spin squeezing in an NMR quadrupolar system using a lyotropic liquid crystal sample  at temperature of $26^\circ{\mathrm C}$, in order to keep the liquid crystalline phase stable.

A spin-squeezed state can be attained by applying an interaction that depends
non--linearly on Cartesian orbital angular momentum components, such that  $\hat{J}_{\mathbf{n}}=\mathbf{n
}\cdot\left( \hat{J}_{x},\hat{J}_{y},\hat{J}_{z}\right) $
perpendicular to the mean spin $\left\langle \hat{J}\right\rangle $ 
(notice that we use $\hat{J}$ to indicate an orbital  angular momentum and $\hat{I}$ to refer to the nuclear spin angular momentum). This
procedure is the one performed by the OAT model \cite
{kitagawa1993,chaudhury2007} or by the two-axis counter-twisting model \cite
{kitagawa1993,fernholz2008}. The OAT model, that will be
addressed in this letter, is characterized by a quadratic term in the
z-component of orbital angular momentum $\kappa\hat{J}_{z}^{2}$, where $
\kappa$ is the strength of this interaction.  One starts from a coherent spin
state $\left\vert j,j\right\rangle _{x}$, which is in
correspondence to a symmetric quasi-probability distribution under the
spherical phase space around the $x$-axis, and then, after a transformation, it appears squeezed in
the $y$-$z$ plane in a rotated (twisted) basis $y^{\prime}$-$z^{\prime}$ 
\cite{fernholz2008}.

To quantify the degree of squeezing we consider the criteria of Ref. \cite{law2001,jin2007PRA,jin2007PRL} developed from Ref. \citep{kitagawa1993}, such that the parameter of squeezing is defined by
 $ \xi=\left( \Delta\hat{J}_{\mathbf{n}}\right) _{\min}/\sqrt{J/2}$,  in order that  $
\left( \Delta\hat{J}_{\mathbf{n}}\right) _{\min}$ represents the
smallest variance of a spin component $\hat{J}_{\mathbf{n}}$ normal to
the mean spin $\left\langle \hat{J}\right\rangle $, specifically, 
\begin{eqnarray}
\xi & = & \frac{\sqrt{\frac{1}{2}C-\frac{1}{2}\sqrt{A^{2}+B^{2}}}}{\sqrt{J/2}} \ < \ 1
\text{, and} \label{ParametroSqueezing} \\
 \alpha_{\xi} &=& \frac{1}{2}\arctan \left( B / A \right)
\text{,}  \label{FaseSqueezing} 
\end{eqnarray}
where $A=\left\langle \hat{J}_{z}^{2}-\hat{J}
_{y}^{2}\right\rangle $, $B=\left\langle \hat{J}_{z}\hat{J}_{y}+
\hat{J}_{y}\hat{J}_{z}\right\rangle $ and $C=\left\langle \hat{J}
_{z}^{2}+\hat{J}_{y}^{2}\right\rangle $ are appropriate combinations of sums and products about   $\hat{J}_{z}$ and $\hat{J}_{y}$, established by the orientation of $\mathbf{n}$, in this case along $x$. $\alpha_{\xi}$ is the \textit{squeezing angle}, which is a geometrical property that characterizes the orientation of the squeezing \cite{jin2007PRA,jin2007PRL}. In addition, it is worth noticing, that the parameter $\xi$\ is a very
useful concept to characterize entanglement \cite
{hald1999,sorensen2001,helmerson2001}, coherent quantum control \cite
{grond2009} besides spin squeezing \cite{wineland1992,vitagliano2011}.

The NMR formalism under any quadrupolar system is based on a nuclear
spin $I>1/2$ and $m=I,I-1,\ldots ,-I$ as its quantization rule. One uses the
laboratory frame representation to set up the Hamiltonian, with basically four kinds of
contributions: the first one is the Zeeman contribution, due to the
interaction of the nuclear magnetic moment $-\hbar \gamma \left( \hat{I}
_{x},\hat{I}_{y},\hat{I}_{z}\right) $  with a strong static magnetic
field $B_{0}$ aligned in the $z$--direction. This first contribution is
expressed by $-\hbar \gamma B_{0}\hat{I}_{z}$, where $ \gamma $ is the gyromagnetic ratio of the nuclear species and $\hbar $ is the reduced Planck's constant. The second one is the
effective quadrupolar term, which arises from the interaction of the
quadrupole moment ($Q$) of the nuclei with the electric field gradient internally present in the sample ($
V_{\alpha ,\beta }$). It is expressed by $\frac{eQ}{4I\left( 2I-1\right) }
\left( V_{zz}\left( 3\hat{I}_{z}^{2}-\hat{I}^{2}\right) +\left(
V_{xx}-V_{yy}\right) \left( \hat{I}_{x}^{2}-\hat{I}_{y}^{2}\right)
\right) $, and the electric field gradient satisfy the Laplace's
equation $\sum_{\alpha }V_{\alpha \alpha }=0$. In an ordered nuclei system,
with an axial symmetry, the condition, $\left\vert V_{xx}\right\vert \approx
\left\vert V_{yy}\right\vert \ll \left\vert V_{zz}\right\vert $, is satisfied. This allows
to simplify the second contribution in the form, $\frac{eQV_{zz}
}{4I\left( 2I-1\right) }\left( 3\hat{I}_{z}^{2}-\hat{I}^{2}\right)$. 
This term will be the generator of the nuclear spin-squeezed
state. The third one is called the radio--frequency (\textit{RF}) term, due
to the interaction of the nuclear magnetic moment with a time dependent external
 magnetic field perturbation $\mathbf{B}_{1}\left( t\right)
=B_{1}\left( \cos \left( \omega _{RF}t+\phi \right) ,\sin \left( \omega
_{RF}t+\phi \right) ,0\right) $, perpendicular to the strong static magnetic
field $B_{0}$. Such that the interaction is represented by $+\hbar \gamma
B_{1}\left( \hat{I}_{x}\cos \left( \omega _{RF}t+\phi \right) +\hat{I}
_{y}\sin \left( \omega _{RF}t+\phi \right) \right) $, where the phase $\phi $
defines its direction on the $x-y$--plane. Finally, the fourth term is due to
contributions from the environment ($\mathcal{H}_{env}$) and represents
effective weak interactions with, among others, nuclear species, electrons,
and field fluctuations \cite{oliveira2007}. In a rotating frame representation, the total
NMR Hamiltonian is described by 
\begin{align}
\mathcal{H}_{NMR}& =-\hbar \left( \omega _{L}-\omega _{RF}\right) \hat{I}
_{z}+\hbar \frac{\omega _{Q}}{6}\left( 3\hat{I}_{z}^{2}-\hat{I}%
^{2}\right)   \notag \\
& +\hbar \omega _{1}\left( \hat{I}_{x}\cos \phi +\hat{I}_{y}\sin \phi
\right) +\mathcal{H}_{env}^{\prime }\text{,}  \label{hamiltonianoRMNQ}
\end{align}%
where $\omega _{Q}=\frac{3eQV_{zz}}{2I\left( 2I-1\right) \hbar }$ means the
quadrupolar coupling, $\omega _{1}=\gamma B_{1}$ represents the \textit{RF}
strength, and $\omega _{L}=\gamma B_{0}$ is the Larmor frequency of 
nuclear species. The coupling parameters of our physical quadrupolar system
satisfy the inequality $\left\vert \omega _{Q}\right\vert \ll \left\vert
\omega _{L}\right\vert $.

Let us set $\omega _{RF}=\omega _{L}$ and $\phi =0$ to transform the NMR
Hamiltonian in order to get the Hamiltonian (1) of Ref. \cite{jin2007PRL}, 
which corresponds to the one-axis twisting model (when $\omega _{1}=0$) of spin
squeezing, after dropping the constant term $-\frac{\hbar \omega _{Q}}{6}
\hat{I}^{2}$. The Hamiltonian for the experimental setup is then,
\begin{equation*}
\mathcal{H}_{NMR}^{s}=\frac{\hbar \omega _{Q}}{2}\hat{I}_{z}^{2}\text{.}
\end{equation*}

Our NMR experimental setting is realized using Cesium nuclei ($^{133}$Cs)
with quadrupolar spin system $I=7/2$ and the dimension of the Hilbert space
is $d=2I+1=8$. A lyotropic liquid crystal sample was prepared using $42.5$
wt \% of Cesium-Pentadecafluoroctanoate (Cs-PFO) and $57.5$ wt \% of
Deuterated water (D$_{2}$O). The experiment was carried out on a Varian 500 MHz spectrometer with a 5mm probe for liquids. The Larmor and Quadrupolar frequencies of  $^{133}$Cs nuclei  are respectively $\omega_{L}/2\pi=65.598$
MHz and $\omega_{Q}/2\pi=7.58$ kHz. The length of the $\pi $-pulse was
calibrated at 26 $\mu $s. The transverse and longitudinal
relaxation times are $T_{2}\approx \ 30$ ms and $T_{1}\approx \ 650
$ ms, respectively. The recycle delay time is 3.5 s.

The formalism to describe a quantum state in an NMR system, we use the density operator
 at thermal equilibrium regime, in which
populations are represented by the Boltzman-Gibbs distribution. The
density operator is denoted by $\rho =\frac{1}{\mathcal{Z}}\hat{1}
+\epsilon \rho _{0}$, where $\epsilon =\omega _{L}\hbar /k_{B}T\mathcal{Z}$
is the polarization value ($\sim 10^{-6}$), $k_{B}$ is the Boltzmann's
constant, $T$ is the room temperature (in our case $26^\circ{\mathrm C}$), $\mathcal{Z}$ the partition function,
and $\rho _{0}=\hat{I}_{z}$, the deviation density matrix \cite{oliveira2007}. The deviation
density matrix is transformed 
by a method adapted from the strongly modulating pulse technique in order to achieve a nuclear spin coherent state, NSCS, the equivalent of the so-called pseudo-pure state \cite{oliveira2007,auccaise2013} in an NMR qubit system, $\left\vert \zeta\left(\theta,\varphi\right)\right\rangle=\sum_{m=-I}^{I} {2I \choose I + m}^{1/2}\cos(\theta/2)^{I-m} \sin(\theta/2)^{I+m} \mathrm{e}^{-i (I+m) \varphi} \left\vert I,m\right\rangle$, where $ \left\vert I,m\right\rangle$ are eigenstates of $\hat{I}_{z}$ with eigenvalue $m$  \cite{auccaise2013,araujo-ferreira2013}.
The density operator changes to $\rho=\left(  \frac{1-\epsilon}{\mathcal{Z}}\right)  \hat{1}+\epsilon\Delta\rho$, such that $\Delta\rho
\equiv\left\vert \zeta\left(  \theta,\varphi\right)  \right\rangle
\left\langle \zeta\left(  \theta,\varphi\right)  \right\vert $ is the
deviation density operator, for any $0\leq\theta\leq\pi$ and $0\leq\varphi
\leq2\pi$ parameters \cite{auccaise2013,supplementary}. 
Specifically, by choosing $\theta_{0}=\pi/2$ and $\varphi_{0}=\pi$ we implement the initial quantum state denoted by $\left\vert \zeta\left(  \pi/2,\pi\right)
\right\rangle $,  such that it is suitable to implement the spin squeezing protocol \cite{law2001,jin2007PRA,jin2007PRL}.  
Following the nuclear spin squeezing protocol, the $\left\vert \zeta \left(
\pi /2,\pi \right) \right\rangle $ evolves according to the operator defined
by $\mathcal{H}_{NMR}^{s}$ during forty-four different time intervals  and time steps of $\tau _{k+1}-\tau _{k}=3\ \mu $s, with $k=0,1,...,44$, where the discrete
 $\tau _{k}\in \left[ 0,132\ \mu \text{s}\right] $. The read out at each
time step of the evolved initial quantum state is performed using a quantum state
tomography process \cite{supplementary}.

A way to quantify the efficiency of the implementation of $\left\vert \zeta\left(  \pi/2,\pi\right)
\right\rangle $ and its time evolution is the concept of the Wigner quasi-probability
distribution function, which is applied to the experimental deviation density matrix
of the tomographed quantum state $\left\vert \zeta \left( \theta
_{k},\varphi _{k}\right) \right\rangle $. So, by its definition, we have \cite
{agarwal1981,benedict1999,sanchez-soto2013}, 
\begin{equation}
W\left(  \theta ,\varphi \right)  =\sqrt{\frac{2I+1}{4\pi}}
{\displaystyle\sum\limits_{K=0}^{2I}}
{\displaystyle\sum\limits_{Q=-K}^{K}}
\varrho_{KQ} \left(   \theta  ,\varphi ; \theta_{k} ,\varphi_{k}\right)   Y_{KQ}\left(  \theta,\varphi\right)
\text{,} \label{WignerDistributionFunction}
\end{equation}
for $\theta \in \left[ 0, \pi \right]$ and $\varphi \in \left[ 0, 2\pi \right]$,  where $ \varrho_{KQ} \left(   \theta  ,\varphi ; \theta_{k} ,\varphi_{k} \right) = \mathtt{Tr}_{ \theta ,\varphi}\left\{  \left\vert \zeta\left(  \theta_{k},\varphi_{k}\right)
\right\rangle \left\langle \zeta\left(  \theta_{k},\varphi_{k}\right)
\right\vert \hat{T}_{KQ}^{\dagger}\right\} $,  $\hat{T}_{KQ} $ are the spherical tensor operators
(or irreducible tensor operators \cite{agarwal1981,benedict1999,sanchez-soto2013}), $Y_{KQ}\left( \theta
,\varphi \right) $ are spherical harmonics.

\begin{figure}[tbp]
  \begin{center}
    $\begin{array}{c@{\hspace{0.03in}}c}
     \multicolumn{1}{l}{\mbox{\bf }} &
     \multicolumn{1}{l}{\mbox{\bf }} \\ [-0.53cm]
     \epsfxsize=0.73in
     \epsffile{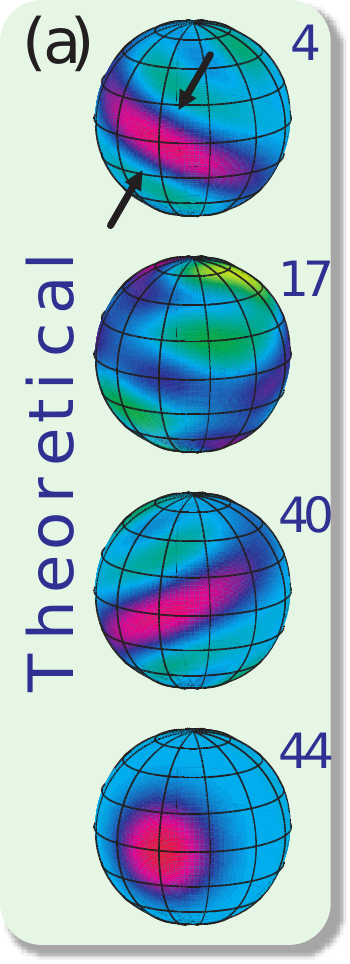} &
     \epsfxsize=2.5in
     \epsffile{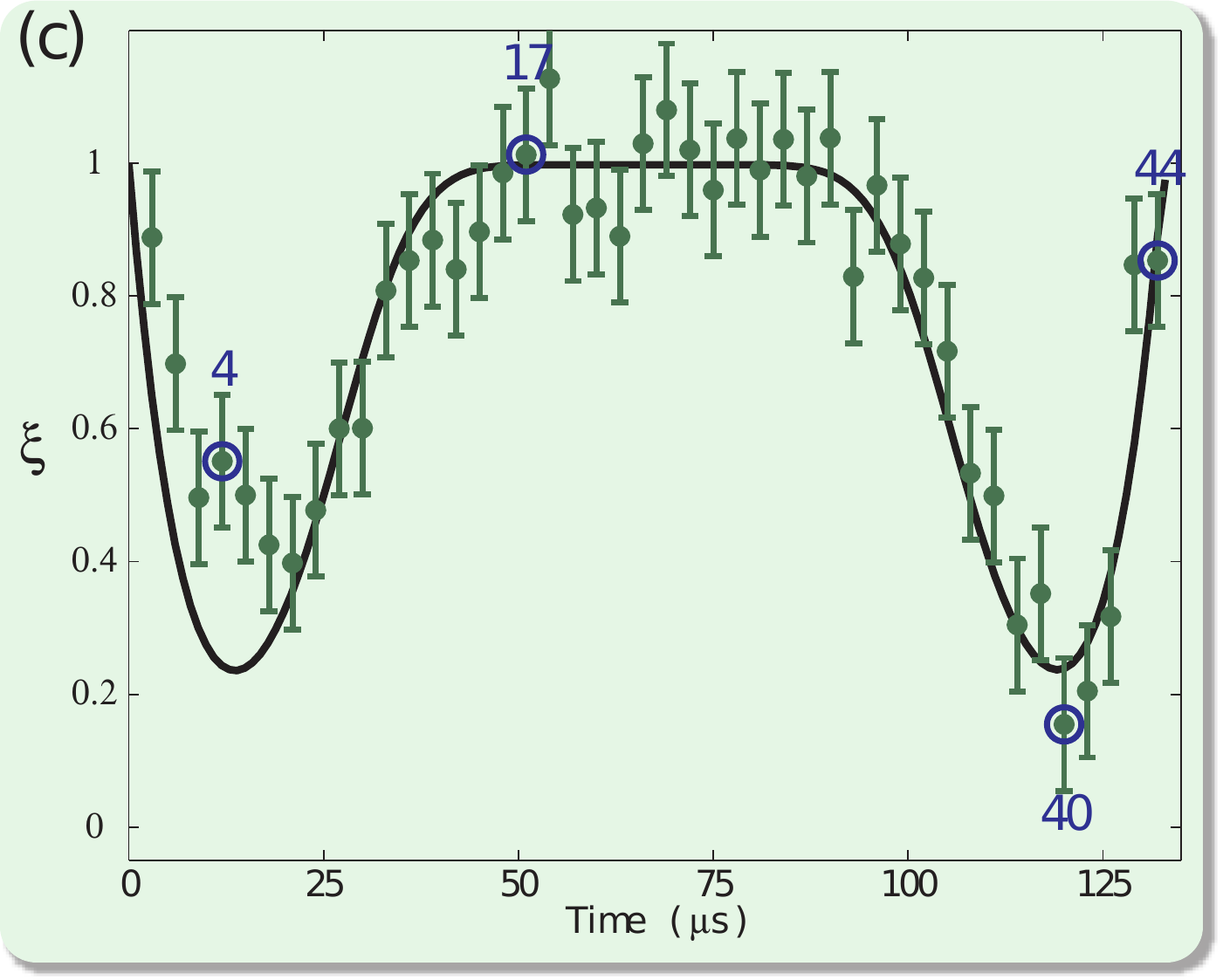} \\ [0.03cm]
     \epsfxsize=0.73in
     \epsffile{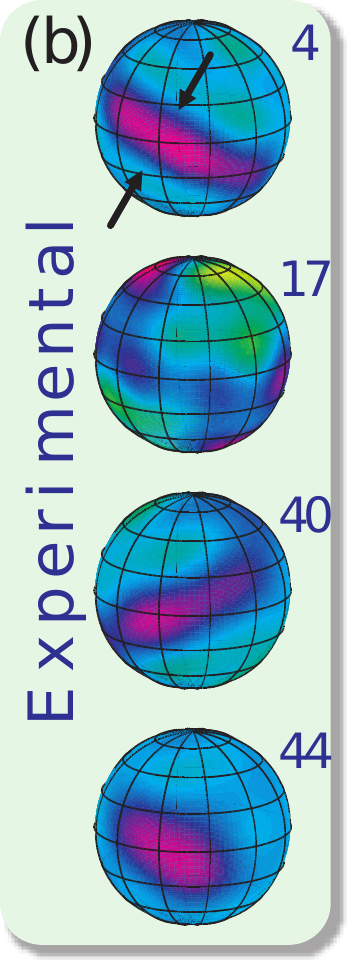} &
     \epsfxsize=2.5in
     \epsffile{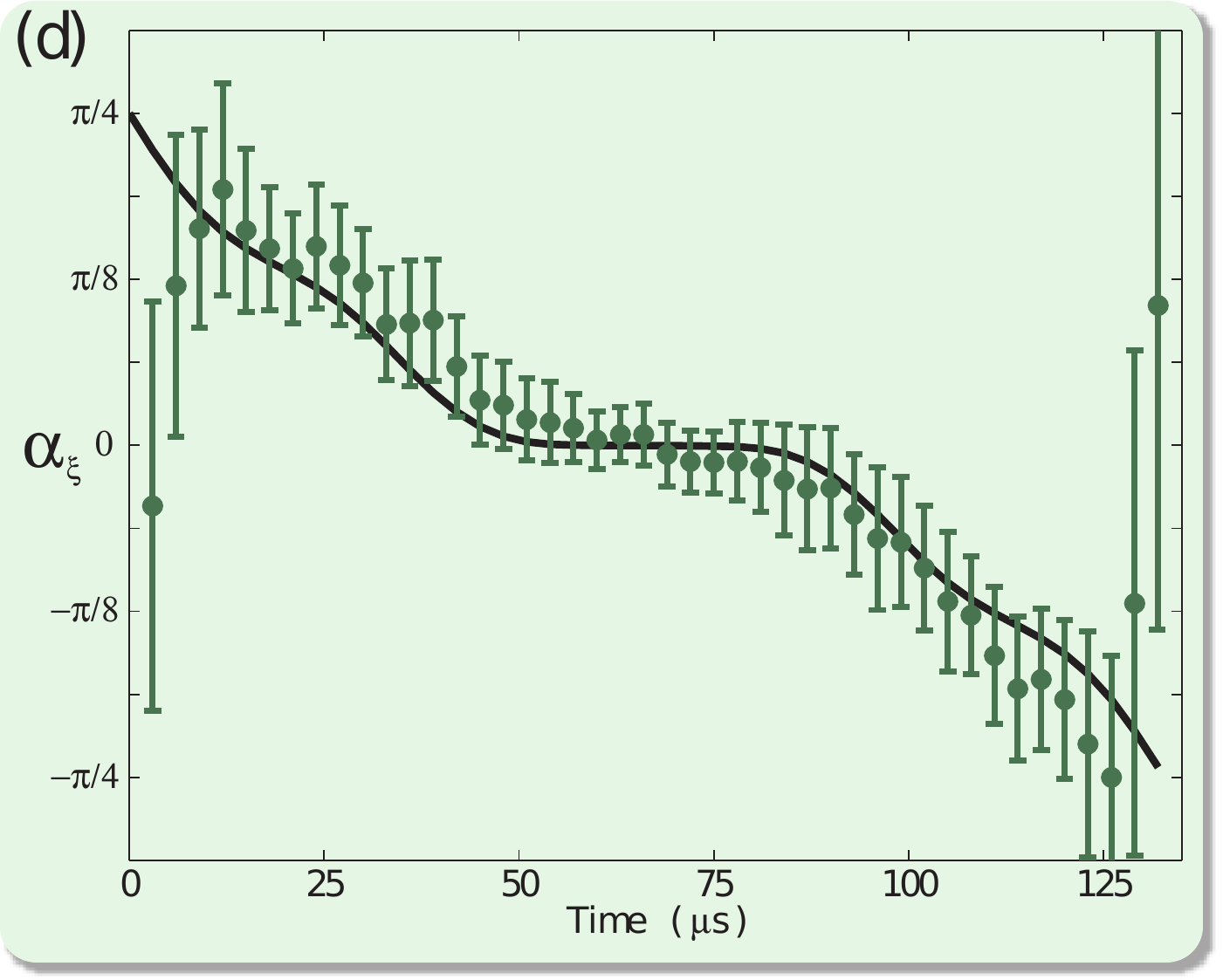} 
     \end{array}$
  \end{center}
  \caption{(Color online) Experimental results of spin squeezing under one-axis twisting model. The NSCS $\left| \protect\zeta\left(\protect \pi/2,\protect\pi\right)\right\rangle$ was evolved under the Hamiltonian $ \mathcal{H}^{s}_{NMR}$. The dynamics of spins was monitored in time steps of $\protect\delta\protect\tau =3 \ \protect\mu$s under the time interval of $\left[0, \protect\nu_{Q}^{-1}\right]$. (a) Theoretical Wigner quasi-probability distribution function computed from a density matrix at time values $\protect\tau_{k}$ where $k=4, 17,  40$ and $44$. (b) Experimental Wigner quasi-probability distribution function calculated from  the tomographed density matrix. (c) Dynamics of the squeezing parameter $\left( \xi \right) $ for theoretical prediction (black solid line), and experimental results (dark green dots). The error bars are discussed in \cite{supplementary}. Blue open circles correspond to analysis of the Wigner quasi-probability distribution function of Fig. \ref{fig:SqueezingSemProtecao}(a) and Fig. \ref{fig:SqueezingSemProtecao}(b).  (d) Dynamics of squeezing angle $\left( \alpha_{\xi} \right)$. }
  \label{fig:SqueezingSemProtecao}
\end{figure}

Next, we analyze the 4-th tomographed quantum state using the
Wigner formalism     
for the theoretical prediction (top of Fig. \ref{fig:SqueezingSemProtecao}(a)) and experimental results (top of Fig. \ref{fig:SqueezingSemProtecao}(b)). We can
observe the qualitative signature of the spin squeezing phenomenon, and we depict the compression of the
distribution probability in the direction denoted by the arrows contained in
the $y$-$z$ plane. Analogously, we show the same phase space at the 17-th time
step,  where the squeezing effect
is attenuated, but at the 40-th time step the squeezing effect is
recovered. Finally, at the 44-th time step we can observe that the next
squeezing cycle due to the shape of its distribution probability has completed the cycle, and started a new one. From that set of figures we observe a correspondence between a 
description following the theoretical development of the matrix operators and a free evolution of the nuclear spins monitored by a quantum state tomography procedure.

We also investigate the dynamics of the squeezing parameter ($\xi $)
computed by Eq. (\ref{ParametroSqueezing}), such that the system evolves under spin squeezing scheme as shown in Fig. \ref
{fig:SqueezingSemProtecao}(c). Theoretical results (black solid line) are generated
transforming a theoretical initial NSCS $\left\vert \zeta \left( \pi
/2,\pi \right) \right\rangle $ under the evolution
operator that depends on the Hamiltonian $\mathcal{H}_{NMR}^{s}$ 
 using numerical recipes for any 
 $\tau \in \left[ 0,132\ \mu \text{s}\right] $. Experimental results
(dark green dots) are computed using forty four tomographed deviation
density matrix, such that $\tau _{k}\in \left[ 0,132\ \mu \text{s}\right] $.
One can observe that the evolution of the spin squeezing has a periodical behavior,
which depends on the inverse of the quadrupolar frequency, $\nu
_{Q}^{-1}=132\ \mu $s, and that it matches with the choice of the time window used to monitor the spin system. The error bars for each experimental dots represents an 
error of \ $\sim$ 10 \%  (see \cite{supplementary} for a detailed discussion).

Similarly, we analyse the evolution of the squeezing angle, Eq. (\ref{FaseSqueezing}), and we show it in Fig. \ref{fig:SqueezingSemProtecao}(d). The theoretical prediction (black solid line) starts at $\pi/4$ value and changes continuously until  -$\pi/4$, which corresponds to the end of the periodical behaviour of the spin system and starts a new one. Precisely at this time value there is a discontinuity of the squeezing angle, that switches from a negative value to positive value. Experimental data (dark green dots) follow closely the solid line computed by the theoretical procedure. We note that the two initial and two final data points do not obey the theoretical curve, this is due simply because the loss of accuracy in the computation of the rate between $B$ and $A$ is reflected through a time delay of a few microseconds to match the transition and follow  the theoretical prediction.

It is worth noticing that the present development of the spin squeezing process  is
different from that reported in solution NMR  \cite{sinha2003}. Indeed, there are two main differences: the first one concerns the kind of interaction used in solution NMR \cite{sinha2003} to produce the squeezing. In the solution NMR procedure, the squeezing is brought off by combining free evolutions of nuclear spins under the effect of $J$--couplings and radio frequency pulses, in order to mimic the effect of the nonlinear Hamiltonian.  In our case, we realized the whole spin squeezed state process by monitoring free evolutions of a nuclear quadrupolar spin system. We can say that in a natural way the quadrupolar nuclei generates the spin squeezing process, because we explore inherent physical properties of the nuclei such as the quadrupole moment and electric field gradient. The second difference is related to the number of nuclear spins of a molecule used to produce the spin squeezing. For solution NMR procedure one uses an ensemble of molecules in which $N$ nuclear spins (one half) interact to achieve the task. In our development we use an ensemble of Cs-PFO molecules in which one quadrupolar nuclei is used to attain our objective.

To conclude, we accomplished, in a liquid-crystal NMR quadrupolar nuclear spin system, an experimental characterization of a spin squeezing process via a one-axis twisting model \cite
{kitagawa1993,chaudhury2007} at a regime of low dimensionality. The theoretical framework established in the atomic physics formalism for symmetrical traps of two--mode BECs is applied to the case of nuclear spin systems respecting their algebraic structure and their commutation rules. The spin squeezing is realized as a result of inherent physical properties of nuclei, and not by the interaction between spins. Although our experimental setup runs at room temperature levels, the pure part of the density matrix, which is proportional to $\epsilon$, holds the quantum behaviour of the nuclear spin system and indicates this squeezing phenomena. The present implementation opens many new possibilities for future applications, for example, by exploring other properties of quadrupolar nuclei, such as: studies on other anisotropies of nuclear quadrupolar systems compatible with the two-axis counter-twisting model \cite{kitagawa1993,fernholz2008}; studies of a prolate, or an oblate, charge distribution around the nuclei compatible with negative, or positive, strength of the nonlinear interaction, or also, the prospect of extending this experimental liquid crystal scheme to an ordered solid state regime.

\begin{acknowledgements}
The authors acknowledge financial support from CNPq, CAPES, FAPESP, and FAPERJ. We thank B. Juli\'a-D\'{\i}az for many invaluable suggestions to our manuscript. R. A. acknowledge M. H. Y. Moussa by the fruitful discussions about atomic coherent states and spin squeezed states. We also acknowledge P. Judeinstein for Cs--PFO samples. This work was performed as part of the Brazilian National Institute of Science and Technology for Quantum Information (INCT-IQ).
\end{acknowledgements}

\end{document}


\title{Supplemental Material: Spin squeezing in an ensemble of quadrupolar NMR system}

\author{R. Auccaise} \email{raestrada@uepg.br}
\affiliation{
Departamento de F\'{\i}sica, Universidade Estadual de Ponta Grossa, Av. Carlos Cavalcanti, 4748, 84030-900 Ponta Grossa, Paran\'{a}, Brazil}

\author{A. G. Araujo-Ferreira}
\affiliation{
Instituto de F\'{\i}sica de S\~{a}o Carlos, Universidade de S\~{a}o Paulo, Caixa Postal 369, 13560-970 S\~{a}o Carlos, S\~{a}o Paulo, Brazil}

\author{R. S. Sarthour}
\affiliation{
Centro Brasileiro de Pesquisas F\'{\i}sicas, Rua Dr. Xavier Sigaud 150, 22290-180 Rio de Janeiro, Rio de Janeiro, Brazil}

\author{I. S. Oliveira}
\affiliation{
Centro Brasileiro de Pesquisas F\'{\i}sicas, Rua Dr. Xavier Sigaud 150, 22290-180 Rio de Janeiro, Rio de Janeiro, Brazil}

\author{T. J. Bonagamba}
\affiliation{
Instituto de F\'{\i}sica de S\~{a}o Carlos, Universidade de S\~{a}o Paulo, Caixa Postal 369, 13560-970 S\~{a}o Carlos, S\~{a}o Paulo, Brazil}

\author{I. Roditi} \email{roditi@cbpf.br}
\affiliation{
Centro Brasileiro de Pesquisas F\'{\i}sicas, Rua Dr. Xavier Sigaud 150, 22290-180 Rio de Janeiro, Rio de Janeiro, Brazil}

\maketitle


\section{\textbf{Details on experimental procedures\label{sec:level1}}}

\begin{figure}[ptb]
\includegraphics[width=0.48	\textwidth]{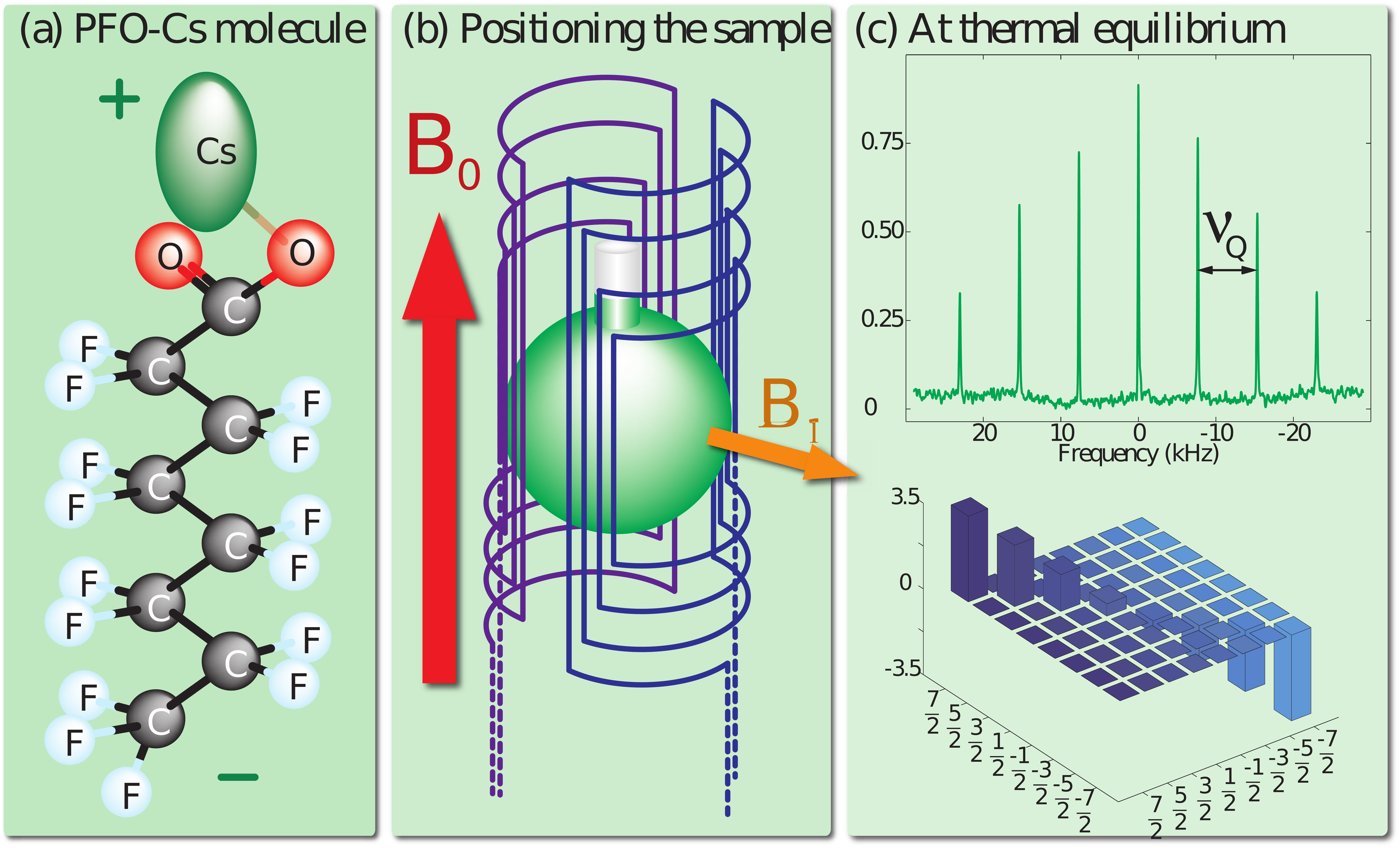}
\caption{(Color online) (a) 
Cesium-Pentadecafluoroctanoate (Cs-PFO) molecule. (b) NMR probe scheme of
the liquid crystal sample placed in a glass bubble. (c) At the top, 
 the typical equilibrium spectrum of $^{133}$Cs nuclei, such that the splitting of the spectral lines corresponds to the quadrupolar frequency $\nu_{Q}$. At the bottom,
a bar chart of the experimental elements of the deviation density matrix
at thermal equilibrium state.}
\label{fig:NMRscheme}
\end{figure}

The present Nuclear Magnetic Resonance (RMN) development is performed in a sample of Lyotropic liquid crystal prepared with Cesium-Pentadecafluoroctanoate (Cs-PFO) molecules and solved in deuterated water\cite{boden1993,jolley2002}. In Fig. \ref{fig:NMRscheme}(a) it is sketched a graphical representation of the molecular structure of the Cs-PFO. The liquid crystal sample was placed in a glass bubble of 4.5 mm external diameter enclosed in a 5 mm NMR tube, where the glass bubble with the sample was placed in the middle of the coil probe as displayed in Fig. \ref{fig:NMRscheme}(b). The target nuclei are $^{133}$Cs such that the characteristic spectra, after a $\pi/2$-pulse, is shown at the top of Fig. \ref{fig:NMRscheme}(c), where its quadrupolar coupling strength corresponds with the splitting of its spectral lines labelled by $\nu_{Q}$.

NMR spin systems, at a room temperature regime, are described by their almost maximum mixture states, such that the associated density matrix has the following form\cite{oliveira2007} 
\begin{equation}
\rho \approx \frac{1}{{\mathcal{Z}}}\hat{1}+\frac{\beta {\hbar \omega _{L}%
}}{{\mathcal{Z}}}\hat{I}_{z}\text{,}  \label{densitymatrix}
\end{equation}
where $\beta =1/k_{B}T$ and ${\mathcal{Z}}={\mathtt{Tr}\left[ e^{\left( -\beta \mathcal{H}_{NMR}\right) }\right] }$ is the partition function, $T$
the room temperature   at $ 26^{\circ }$ (in order to keep the liquid crystalline phase stable), $k_{B}$ the Boltzmann's constant, ${\hbar }$ the reduced Planck's constant, and ${\omega _{L}}$ the Larmor frequency of the nuclei. In this particular case, for $^{133}$Cs nuclei at 11.7 Tesla, the polarization factor is $\epsilon =\frac{\beta {\hbar \omega _{L}}}{{\mathcal{Z}}}=1.3\times 10^{-6}$ which is a slight deviation from the normalized identity matrix. Therefore $\hat{I}_{z}$ represents the so called deviation density matrix, which we denoted by $\rho _{0}$. The experimental deviation density matrix was reconstructed (bottom of Fig. \ref{fig:NMRscheme}(c)) following a quantum state tomography procedure \cite{teles2007}.

\section{\textbf{Details on spin squeezing procedures\label{sec:ssp}}}

\begin{figure}[ptb]
\includegraphics[width=0.48		\textwidth]{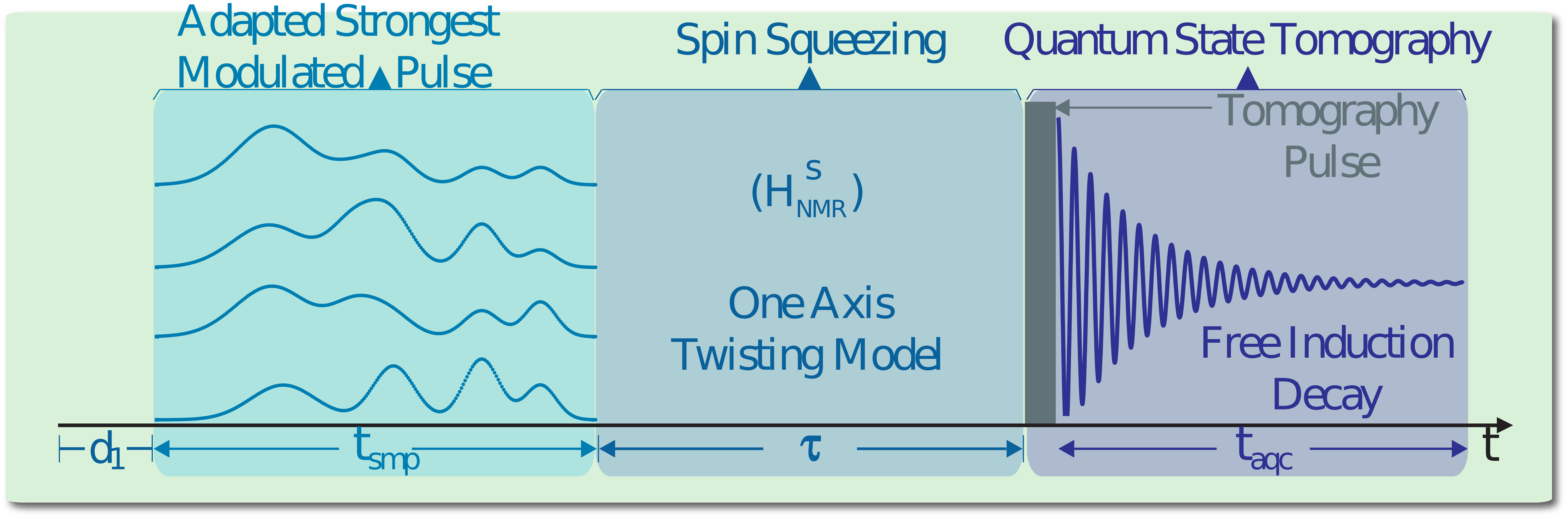} \caption{(Color online) Pulse sequence used to implement
experimentally the nuclear spin-squeezed state. It is divided in three steps. In the first one, we prepare the quantum state $\left|\zeta\left(  \pi/2,\pi\right)  \right\rangle $ using an adapted strongly modulated pulse technique. In the second one, the nuclear spin-squeezed regime is achieved by a free evolution of the quantum spin system under the Hamiltonian $\mathcal{H}_{NMR}^{s}$
explained in the main text. Finally we reconstruct the quantum state using the quantum state tomography procedure.}
\label{fig:PulseSequence}
\end{figure}

The Spin squeezing experimental implementation may be illustrated by the pulse sequence depicted in Fig. \ref{fig:PulseSequence}, which is divided into three stages: the first one corresponds to the application of an adapted strongly modulated pulse in order to transform the deviation density matrix $\rho _{0}$ into an initial nuclear spin coherent state, (NSCS), which is an equivalent description of the so-called pseudo-pure state in an NMR system \cite{oliveira2007,auccaise2013} denoted by $\Delta \rho \left( 0\right)  \equiv\left\vert \zeta\left(  \theta,\varphi\right)  \right\rangle \left\langle \zeta\left(  \theta,\varphi\right)  \right\vert $, which we name as nuclear spin coherent state (NSCS). Next, we perform the spin squeezing scheme (via the one axis twisting model Hamiltonian) transforming $\Delta \rho \left( 0\right) $ into $\Delta \rho \left( \tau \right) $, and finally we reconstruct the transformed quantum state using the quantum state tomography procedure. 

Independently of those stages, a recycle time delay ($d_{1}$) is introduced before the first stage of the pulse sequence, to guarantee that the spins system returns to the thermal equilibrium state after any spin manipulation, only then it is possible to repeat the experimental protocol. Those three stages are described briefly, as follows.

\subsection{\textbf{Adapted Strongly Modulated Pulse\label{sec:level1A}}}

Employing appropriate spin rotations -- radio frequency pulses and free evolutions -- it is possible to produce a pseudo pure state \cite{gershenfeld1997,cory1997}, which is also possible using soft pulses as is the case in a strongly modulated pulse  \cite{fortunato2002}. More precisely, the adapted strongly modulated pulse technique is a sequence of $N$  radiofrequency pulses at fixed time intervals, such that each one has different values of amplitudes and phases. In this case we impose a particular feature, which needs to be smoothed by a modulation of a set of harmonic functions  and multiplied by gaussian functions at the beginning, in order to raise the amplitude from null intensity, and at the end to decrease the amplitude until close to zero, without abrupt changes. For this reason such procedure is called an \textit{adapted} strongly modulated pulse. We construct it with the purpose of eliminating undesired transitions during the implementation of the radio frequency pulse. Those values are chosen judiciously in order to achieve a specific transformation to produce a target state (which is known a priori). The procedure works as follows: (i) Initially a random choice of values, for the amplitude and phase, defines a general rotation matrix (total of $2N$ pulse parameters), which transforms the deviation density matrix into a new one,  that depends on the pulse parameters. By summing up an established number $\eta$ of these transformed states (usually one takes  $\eta=4$) an average state is obtained. (ii) Next, comparing it with the target state it is possible to build a fidelity function. So, an optimization algorithm (in our case SIMPLEX Nelder-Mead \cite{nelder1965}) is applied to obtain the $2\eta N$ pulse parameters that maximize the fidelity function. (iii) Once the optimal pulse parameters are defined, the pulse sequence comprised by $N$ pulses with amplitudes and phases is implemented at each $\eta$-stage in the NMR spectrometer producing the desired state.  In this way, the thermal state (Eq. (\ref{densitymatrix})) can be transformed into a state of the form
\begin{equation}
\rho \approx \left( \frac{1-\epsilon}{\mathcal{Z}} \right) \hat{1}%
+\epsilon \Delta\rho,
\label{pseudopurestate}
\end{equation}
where $\Delta\rho $ represents the NSCS, having the form of a pure state density matrix
with unitary trace.

In Fig. \ref{fig:PulseSequence} we illustrate the application of the adapted strongly modulated pulse for  $N{=}256$ and $t_{smp}{=}2\nu_{Q}^{-1} = 264 \mu$s ( $\nu_{Q}{=}\omega_{Q}/2\pi$ is the quadrupolar frequency) to produce the initial quantum state  $\Delta\rho{=}\left\vert \zeta\left( \theta,\varphi\right)  \right\rangle
\left\langle \zeta\left(  \theta,\varphi\right)  \right\vert $ with angular parameters $\theta{=}\pi /2$ and $\varphi{=}\pi$.

\begin{figure}[ptb]
\includegraphics[width=0.48	\textwidth]{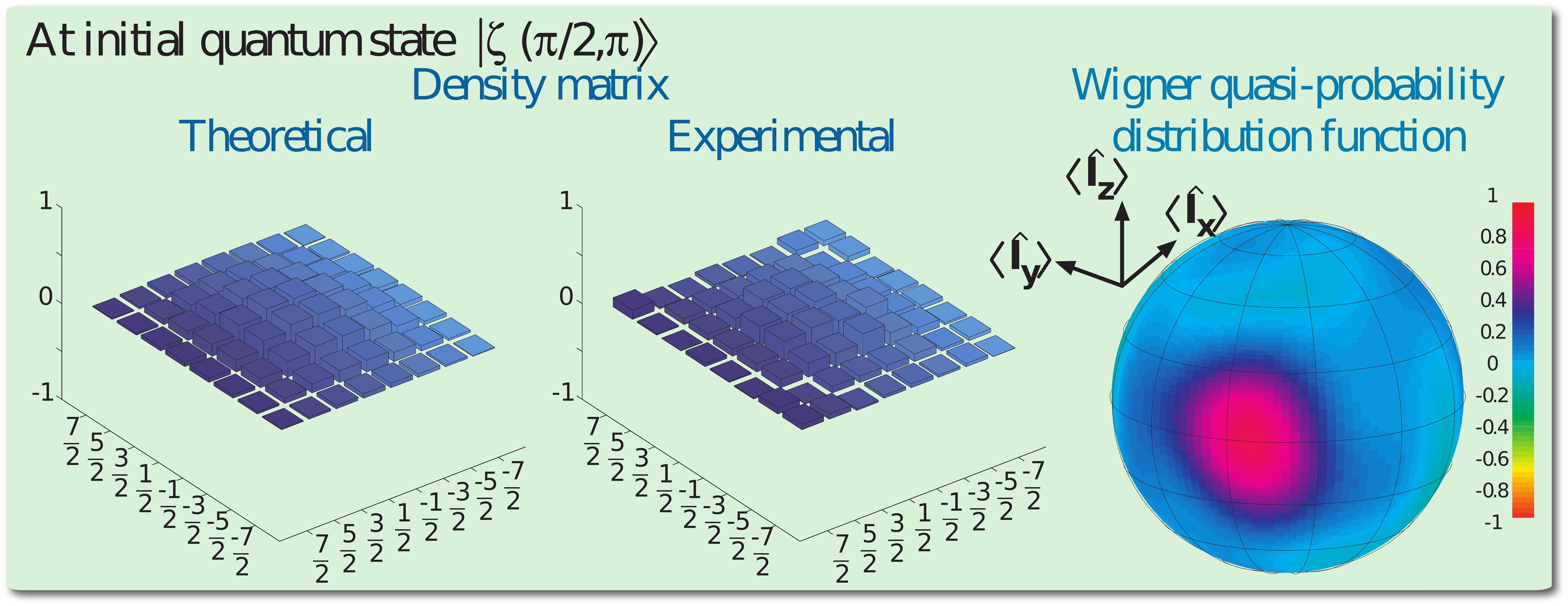}
\caption{(Color online)  At left (middle), we present a bar
chart of the theoretical (experimental) elements of the density matrix
of the quantum state $\left\vert \protect\zeta\left( \protect\pi/2,\protect
\pi\right) \right\rangle $. At right, under the spherical phase space
representation, we show the Wigner quasi-probability
distribution function, Eq. (4) of main text, computed from
the experimental quantum state $\left\vert \protect\zeta\left( \protect\pi/2,
\protect\pi\right) \right\rangle $. The intensity of $W\left(  \theta,\varphi\right) $
 is encoded by the colour bar at right of the sphere .}
\label{fig:NuclearSpinCoherentStateExperimental}
\end{figure}

\subsection{\textbf{The one-axis twisting model in NMR\label{sec:level1B}}}

The one-axis twisting model depends non-linearly on the z-component of the orbital angular momentum operator, $\kappa\hat{J}_{z}^{2}$, in the ultra cold atom description, where $\kappa$ represents the strength of interaction between particles.

On the other hand, the NMR Hamiltonian of a quadrupolar system in the rotating frame representation (see Eq. (1) of the main text) needs to be reduced to an analogous non-linear dependance of the atom description. In an experimental setup, one has to turn off the radio frequency transmitter which means that $\omega_{1}=0$, and the transmitter offset needs to be synchronized to the frequency of the central line of the spectra,see top of Fig. \ref{fig:NMRscheme}(c). The term $\hbar\frac{\omega_{Q}}{6}\hat{I}^{2}$, which is proportional to the identity matrix, does not produce any effect on the quantum state. The effects coming from environment contributions ($\mathcal{H}^{\prime}_{env}$) are weak enough such that we can ignore them in our quantum description.

In this way, it is possible to describe the NMR Hamiltonian at the rotating frame such as
\begin{align}
\mathcal{H}_{NMR}^{s}  &  =\hbar\frac{\omega_{Q}}{2}   \hat{I}_{z}^{2}%
  \text{,} \label{hamiltonianoRMNQ}%
\end{align}

In Fig. \ref{fig:PulseSequence} we present how we carry out the evolution of the one-axis twisting model under the Hamiltonian of Eq. (\ref{hamiltonianoRMNQ}). During a time $\tau$ the implemented  $\left\vert \zeta\left(  \pi/2,\pi\right)  \right\rangle
\left\langle \zeta\left(  \pi/2,\pi\right)  \right\vert $ is transformed, performing the squeezing effect as described in the main text.

It is important to note that we ignore relaxation processes, because the temporal window in which we monitor the evolution of spin squeezing (132$\mu$s) is several orders of magnitude smaller than the transversal relaxation time (30 ms).

\subsection{\textbf{The quantum state tomography procedure\label{sec:level1C}}}

The read out procedure of the implemented or evolved quantum states, in $^{133}$Cs nuclei on a lyotropic liquid crystal sample, is the quantum state tomography using global rotations \cite{teles2007}. In this tomography procedure we implement the following three conditions: (i) an appropriate phase cycling of the receiver, (ii) a precise rotation of the nuclear spin which are described through the irreducible tensor formalism, and (iii) a compatible number of scans.  All the procedures mentioned above are intended to detect a selected order of quantum coherence \cite{araujo-ferreira2012}.

In Fig. \ref{fig:PulseSequence} we show the implemented quantum state tomography procedure as a final stage of the experimental protocol of the nuclear spin-squeezed state process. The tomography pulse carries information about the spin rotation, being usually shorter than  a $\pi/2$-pulse, and the free induction decay carries the appropriate phase cycling in which the receiver is turned on during a 300 ms window of time.

\section{\textbf{Experimental nuclear spin-squeezed state results\label{sec:level2}}}

The primary aim of the present development is the experimental implementation of spin squeezing using a lyotropic liquid crystal system. To compute the experimental data, we repeated the implementation at five \textit{different experiments}, performing a statistics.  This procedure is realized as follows: Using the adapted strongly modulated pulse technique a set of pulse parameters were optimized in order to initialize the desired NSCS $\left\vert \zeta\left(  \pi/2,\pi\right)
\right\rangle $. Next, it is transformed by the spin squeezing protocol, and finally the read out of the quantum state is done by the quantum state tomography procedure, as explained in  Sections \ref{sec:level1A}, \ref{sec:level1B}, \ref{sec:level1C} (see Fig. \ref{fig:PulseSequence}). The data generated at the experiment are the forty four tomography density matrices plus one density matrix that corresponds to the initial quantum state.

Again, we run the adapted strongly modulated pulse technique, but with a different set of pulse parameters, however producing the same  NSCS $\left\vert \zeta\left(  \pi/2,\pi\right) \right\rangle $. After, this state is transformed by the spin squeezing protocol, and finally the tomography quantum state procedure is performed. On this second procedure other forty five density matrices were generated.

We repeated three more experimental runs following the same protocol and keeping constant many physical parameters as room temperature, quadrupolar coupling, recycle time delays, transmitter offsets, $\tau_{i}$-time steps. We only changed the set of pulse parameters generated by the adapted strongly modulated pulse technique.

So, from those five different experimental runs we get five sets, of forty five density matrices. Now, we compute the mean density matrix over the first one density matrix of each set of matrices. For this initial   NSCS $\left\vert \zeta\left(  \pi/2,\pi\right) \right\rangle $, the result we obtained is shown in the bar chart at the middle of Fig. \ref{fig:NuclearSpinCoherentStateExperimental}, and for comparison we display, at left of Fig.  \ref{fig:NuclearSpinCoherentStateExperimental}, the theoretical elements of its density matrix.

Similarly, this procedure is done for the other forty four evolved density matrices, those matrices are not displayed.

On the other hand, to have a qualitative characterization of the initial tomographed quantum state  $\left\vert \zeta\left(  \pi/2,\pi\right) \right\rangle $ we apply the Wigner-quasi probability distribution function definition (see Eq. 4 of the main text),  and its result is depicted using the spherical phase space representation at the right of Fig. \ref{fig:NuclearSpinCoherentStateExperimental}, the radius of the sphere is normalized to one in our description. We can visualize a symmetric distribution of probability strength around the negative $x$-direction, which is a characteristic of the NSCS $\left\vert \zeta \left( \pi /2,\pi \right) \right\rangle $. Also, we can observe signatures of small variations between small negative and positive probabilities due to imperfections of the measured density matrix, which will be addressed to experimental errors explained in section \ref{sec:level3}.

An analogous procedure for the experimental results shown in Fig. 1 of the main text were performed, as detailed above.

\section{\textbf{Sources of experimental error}\label{sec:level3}}

The description of our experimental implementation on spin squeezing depends on the accuracy of methods to control and to detect the nuclear spin system. Here we describe the main sources of experimental error which may alter its precision.

\begin{figure}[ptb]
\includegraphics[width=0.48	\textwidth]{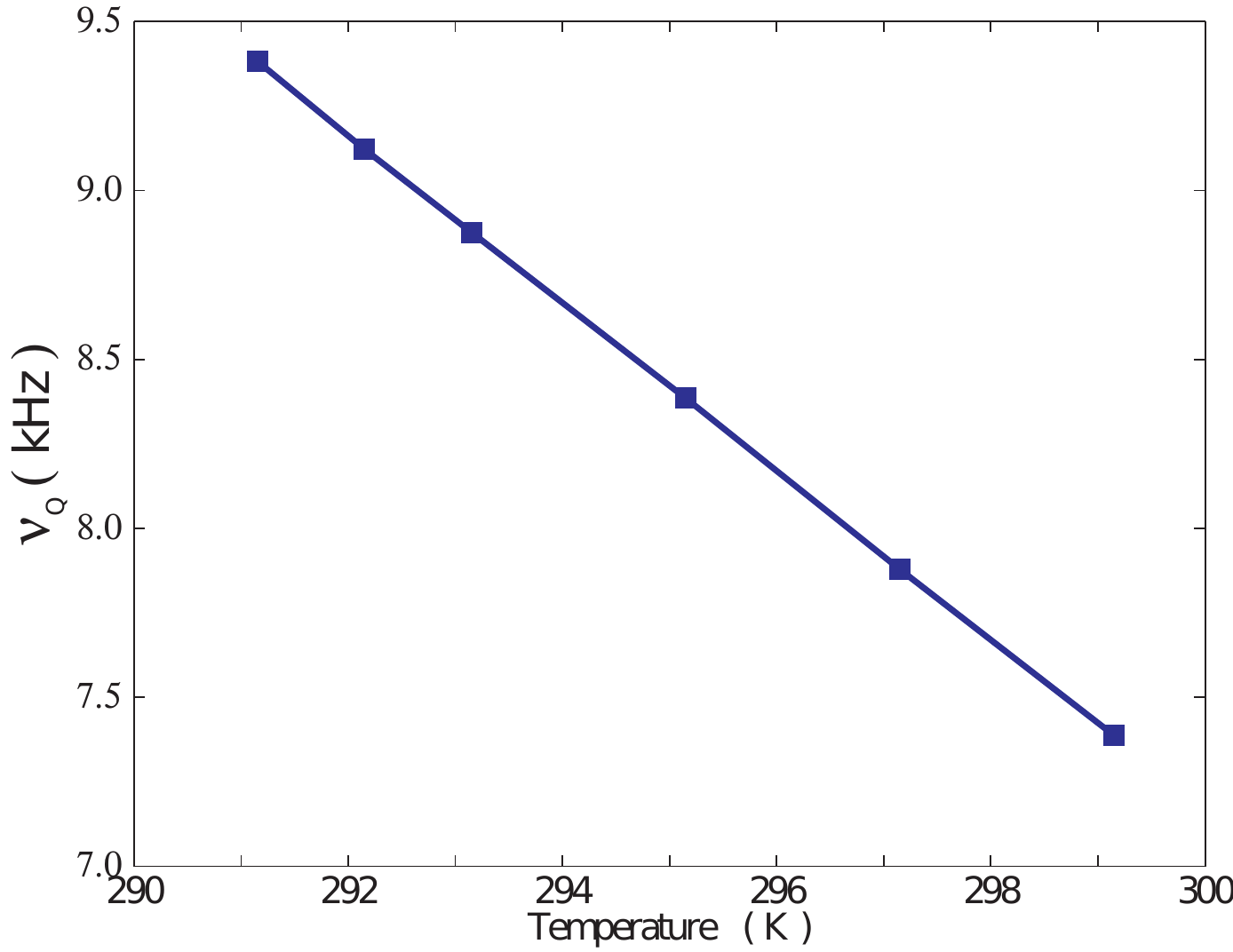} \caption{(Color online) We show experimental data of quadrupolar frequency ($\nu_{Q}$) against temperature. Solid line joins the experimental symbols. The average slope is -250 Hz/K.}
\label{fig:QuadCouplingVSTemperature}
\end{figure}

\subsection{Source of errors for quantum state tomography}
Quantum state tomography procedures are efficient protocols to reconstruct the density matrix in an optimal experimental configuration. In our experimental set up, it corresponds with a pure spin rotation that depends uniquely on a radio frequency term of the NMR Hamiltonian (see Eq. (3) of main text). In soft-matter and solid-state physical systems, this condition is not always fulfilled. Frequently the external radio frequency strength ($\omega_{1}$) is much greater than interaction strength between particles (quadrupolar frequency ($\omega_{Q}$) in our case), and also short lengths of pulses. In the present work those physical parameters are quantified by the following values: $\omega_{1} /2\pi = 19$ kHz,  $\omega_{Q} /2\pi = 7.58$ kHz, and 26 $\mu$s of a $\pi$-pulse. At this configuration of parameters, the discussion about the experimental accuracy proceeds as follows: let us assume that the tomography pulse $U\left( t\right) $ depends on Hamiltonian of Eq. (3) of the main text, such that at resonance $    \omega _{L}=\omega _{RF} $, it  is expressed by the  operator $U\left( t\right)  = \exp\left[ -i \omega _{1}t  \left( \hat{I}_{x}\cos \phi +\hat{I}_{y}\sin \phi \right)   -i \frac{\omega _{Q}t}{2} \hat{I}_{z}^{2}   \right]   $, where $t$ is the time of the tomography pulse. By the linear contribution of the pulse operator, $\omega _{1}t $  could represent any nutation angle of table I on reference  \cite{araujo-ferreira2012}. At the non-linear contribution, $\frac{\omega _{Q}t}{2}$ represents an azimuthal evolution of the spin system. Both of them generate a common dynamic of the nuclear spin during the  tomography pulse, where the azimuthal contribution diminishes its accuracy. To have an idea about the loss of precision that introduces the azimuthal rotation, we compare the operator $U\left( t\right) $ with an ideal pulse  $V\left( t\right)  = \exp\left[ -i \omega _{1}t  \left( \hat{I}_{x}\cos \phi +\hat{I}_{y}\sin \phi
\right)   
  \right]   $. To accomplish this task, we use the lowest angle of nutation to detect the zero order coherence from the Table I of reference   \cite{araujo-ferreira2012}. If we consider the above physical parameters, then the time of the pulse is  $t=2.2 \ \mu$s. So, we define the orientation of the pulse along the positive x-axis ($\phi=0$), and we compute the fidelity\citep{fortunato2002} between two operators  $\mathcal{A} =U\left( t\right)$ and $\mathcal{B}=V\left( t\right)$ defined by Eq. (\ref{FidelidadeUV}). The value of $F = 97.2$ \% represents the degree of similarity between $U$ and $V$ or 2.8 \% of error when we use the operator $U$ as a tomography pulse.
\begin{equation}
F=\frac{{\mathtt{Tr}}\left\{ \hat{\mathcal{A}}\cdot \hat{\mathcal{B}}^{\dagger}\right\} }{\sqrt{{%
\mathtt{Tr}}\left\{ \hat{\mathcal{A}}\cdot \hat{\mathcal{A}}^{\dagger} \right\} {\mathtt{Tr}}%
\left\{ \hat{\mathcal{B}}\cdot \hat{\mathcal{B}}^{\dagger}\right\} }}\text{.}
\label{FidelidadeUV}
\end{equation}
An analogous procedure is done for other pulses of the tomography protocol.

\subsection{Source of errors from variations of temperature}
Among the many physical properties of liquid crystals, the dependance between quadrupolar frequency and temperature is an important characteristic of this kind of soft material. So, fluctuations in temperature induce also fluctuations in quadrupolar frequency. To put in evidence how much this property diminishes the accuracy of our implementation, we performed an experiment of direct detection of NMR signal (it means that a recicle delay is considered, next we apply a $\pi / 2$-pulse, and finally detect the signal). In Fig. \ref{fig:QuadCouplingVSTemperature},  we present the dependence of quadrupolar frequency under variation of temperature\citep{boden1993,jolley2002}. We observe a linear response proportional to -250 Hz/K. If the accuracy of the temperature controler is $\pm 0.1$ K then we can identify one source of error in our experimental set up of  $\pm 25$ Hz  of the quadrupolar frequency, therefore, it assumes the value  $7580 \pm 25$ Hz. 

\subsection{Source of errors from electronic devices of the spectrometer}
A Spectrometer is an aparatus of high precision, but in some sense its precision is restricted by the  speed of response of electronic devices on it. Particularly in our NMR apparatus, if we want to perform an experiment of direct detection as explained above, the sequence of events are not instantaneous. For an event to take place, it takes at least a minimum time to accomplish it or the \textit{hidden delays} (see reference \cite{ManualVnmrJ} for more details). For example, a radio frequency pulse is divided into three stages: beginning, middle, and end. At the beginning, the electronic device is started such that the phase of the pulse is set at approximately 50 ns, and also the switching of the transmitter gate takes less than 50 ns.  At the middle stage, the time of the radio frequency pulse is executed, which is stablished by the value of the pulse parameter.  At the end, the electronic device is turned off such that the switching time is less than 50 ns. In this case, the beginning and the end stages are considered as hidden delays.  Other examples are:  the minimum time between  turning on the detector and starting to acquire the first data point, or between two data points of the free induction decay, where it takes 200 ns (sampling interval) -- and the elapsed time to change the frequency value from one value to other one, it takes 4 $\mu$s. We mentioned some generators of hidden delays during a pulse sequence but there are other ones \cite{ManualVnmrJ}, such that their total contribution generates a kind of systematic error.

\begin{figure}[ptb]
\includegraphics[width=0.48	\textwidth]{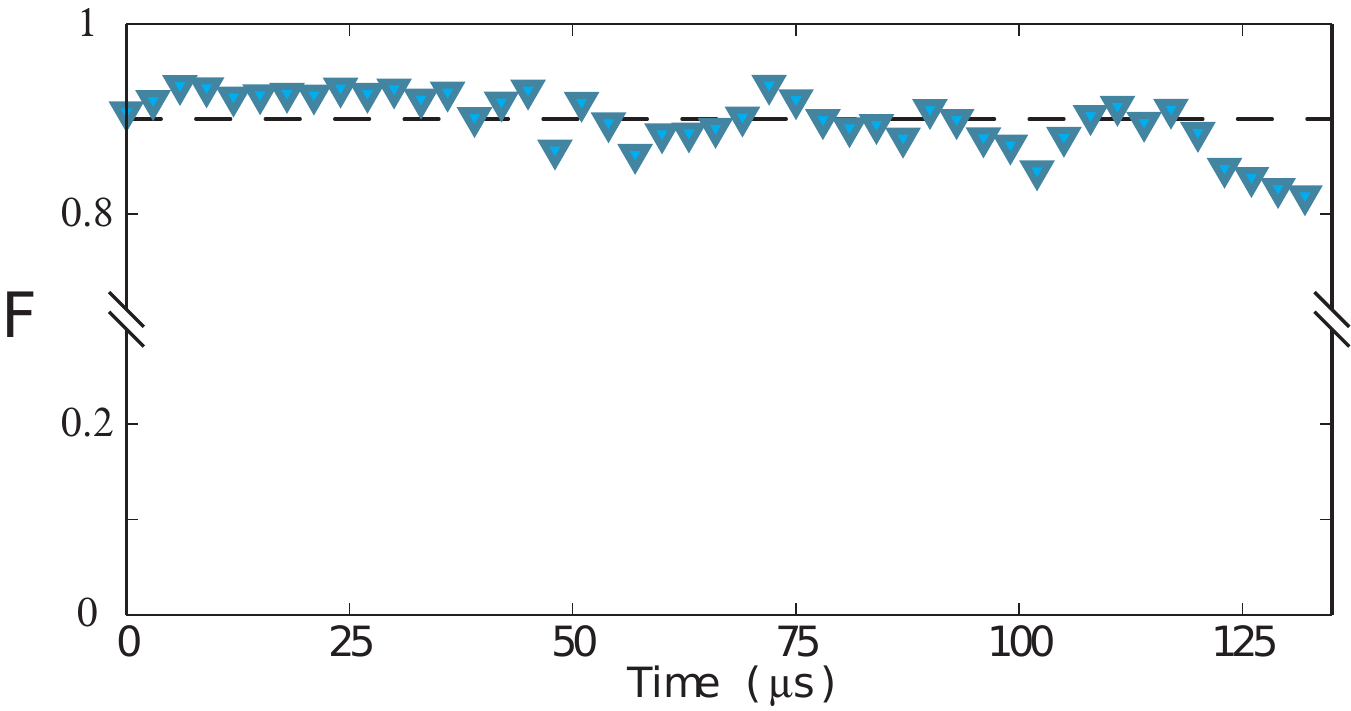} \caption{(Color online) Values of fidelity computed applying Eq. (\ref{FidelidadeUV}) for the fourty five tomographed deviation density matrices and its corresponding theoretical matrices. The dashed line represents an accuracy of 0.9 only as a guide line. }
\label{fig:Fidelidade}
\end{figure}

\bigskip

We have provided here, details concerning the main sources of error on our experimental set up, but not all of them, as there are other possibilities such as inhomogeneities of the magnetic fields or fluctuations of the frequency offsets. In order to quantify all sources of error we use the concept of fidelity \cite{fortunato2002}. Its values are obtained by the comparison between the tomographed deviation density matrix $\rho _{Exp}=\hat{\mathcal{A}}$ and the
theoretical prediction $\rho _{The}=\hat{\mathcal{B}}$,  we quantify the accuracy of our experimental set up by Eq. (\ref{FidelidadeUV}).

In Fig. \ref{fig:Fidelidade} the symbols (blue triangles) present the result computed by Eq. (\ref{FidelidadeUV})  and the dashed line represents {90\%} of the fidelity as a guide line.  So, the generation and control of our experimental implementation is close to {90\%}, or equivalently a {10\%} of error on the computation of mean values of any spin angular momentum operator. This procedure can be extended to the mean values of $A$, $B$, and $C$ of the main text, and provides the error bars of our experimental results of Fig. 1 in the main text.